\documentstyle[12pt]{article}
\def\lsim{\mathrel{\rlap{\lower4pt\hbox{\hskip1pt$\sim$}}
    \raise1pt\hbox{$<$}}}         %less than or approx. symbol
\def\gsim{\mathrel{\rlap{\lower4pt\hbox{\hskip1pt$\sim$}}
    \raise1pt\hbox{$>$}}}         %greater than or approx. symbol

\def\overleftrightarrow#1{\vbox{\ialign{##\crcr
    $\leftrightarrow$\crcr
    \noalign{\kern 1pt\nointerlineskip}
    $\hfil\displaystyle{#1}\hfil$\crcr}}}
\long\def\caption#1#2{{\setbox1=\hbox{#1\quad}\hbox{\copy1%
\vtop{\advance\hsize by -\wd1 \noindent #2}}}}

\newcommand \beq{\begin{eqnarray}}
\newcommand \eeq{\end{eqnarray}}

\newcommand \ps {P(\sigma)}

\newcommand{\bea}{\begin{eqnarray}}
\newcommand{\eea}{\end{eqnarray}}

\begin{document}
\pagenumbering{arabic}
\setcounter{page}{1}
\thispagestyle{empty}
\newcommand{\rdeg}{$^\circ$}
\raggedbottom
\begin{flushright}
FNAL Experiment E781 Note H-671\\
Tel Aviv U. preprint TAUP-2154-94\\
\end{flushright}
\vspace{1.2cm}
\begin{center}
{\bf TWO-MESON AND MULTI-PION FINAL STATES\\ FROM 600 GeV PION INTERACTIONS}
\end{center}
\vspace{1.4cm}
\begin{center}
{\bf
Murray A. Moinester \\
Raymond and Beverly Sackler Faculty of Exact
Sciences,\\
School of Physics,
Tel Aviv University, 69978 Ramat Aviv, Israel\\
E-mail: MURRAY@TAUPHY.TAU.AC.IL\\
20 May 1994}
\end{center}
\vspace{1.4cm}
\date{}
\normalsize
\centerline{\bf {ABSTRACT}}
\setlength{\baselineskip}{0.3in}
\noindent

This report describes the transitions $\pi^- \rightarrow meson_1 + meson_2$
and also $\pi^- \rightarrow  multi-\pi$ for high energy pions interacting with
target nuclei (Z,A). The physics interests are: A)
Nuclear inelastic coherent diffraction cross sections for
pions, for studies of size fluctuations in the pion wave function. B)
Radiative widths of excited meson states, for tests of
vector dominance and quark models. C) Experimental determination of the
$\pi^- + \rho \rightarrow \pi^- + \gamma$ total reaction rate for photon
production above 0.7 GeV, needed for background studies of quark-gluon
plasma formation experiments. D) Investigation of the $\gamma \rightarrow 3
\pi$ vertex in pion pair production by a pion, for a significantly improved
test of the hypothesis of chiral anomalies.  The physics interest and
associated bibliography are summarized here; with reference to the 200-600
GeV beams available at CERN and FNAL. Complementary GEANT simulations and
trigger studies are needed.
\begin{center}
------------------------------------------------------------------------------\\
\end{center}
\medskip
\medskip
\medskip
\newpage
\centerline{\bf {INTRODUCTION}}

There are a number of different physics objectives (A-D, below) for studies
in E781 \cite{russ} and elsewhere
of the reactions 1-8 listed below, reactions involving
mainly pion induced two-meson final states. The discussion here describes
the different objectives separately, and the relevant data needed for each
objective. The objectives are: A)
nuclear inelastic coherent diffraction cross section channels for incident
pions or hadrons, for studies of color fluctuations. B)
Radiative widths of excited Meson States Via Primakoff. C)
determination of the $\pi^- + \rho \rightarrow \pi^- + \gamma$ total
reaction rate for photon production above 0.7 GeV, via Primakoff. D)
Investigation of the Chiral Anomaly $\gamma \rightarrow 3 \pi$ in pion pair
production by a pion, via Primakoff. The Primakoff studies B,C,D are rather
straightforward, the
diffractive studies need to be studied as background to the
Primakoff reactions; but here are discussed for their own merits.
Detailed discussions of items A-D are given below.\\

Examples of the reactions considered are: \\
1) $\pi^-$ + virtual photon $\rightarrow  \pi^-  + \rho$ (Primakoff)\\
2) $\pi^-$ + pomeron $\rightarrow \pi^-  + \rho$ (Diffractive)\\
3) $\pi^-$ + virtual photon $\rightarrow K^-  + K^*$ (Primakoff)\\
4) $\pi^-$ + pomeron $\rightarrow K^-  + K^*$ (Diffractive)\\
5) $\pi^-$ + virtual photon $\rightarrow \pi^-  + \omega$ (Primakoff)\\
6) $\pi^-$ + pomeron $\rightarrow \pi^-  + \omega$ (diffractive,
not allowed by G-parity)\\
7) $\pi^-$ + pomeron $\rightarrow$  L pions  + M kaons \\
8) hadron + pomeron $\rightarrow$ hadron' meson
(for example, proton $\rightarrow$ Lambda K$^+$,\\
or $\pi^- \rightarrow \bar{p} p \pi^-$)

\newpage
\noindent
Some quantum numbers relevant to the above reactions are: \\
    pion: C=+1, G-parity=-1, Isospin=1\\
     rho: C=-1, G=1, I=1\\
   omega: C=-1, G=-1, I=0\\
 pomeron: C=+1, G=1\\
a$_1$(1260): C=+1, G=-1, I=1,      pi-rho decay dominant\\
b$_1$(1235): C=-1, G=+1, I=1,   pi-omega decay dominant\\
  gamma: G=-1,I=0 or G=+1,I=1\\

We distinguish between Primakoff production of a final state $\pi\rho$
configuration, via the reaction $\pi$ + virtual photon $\rightarrow \pi +
\rho$; and a diffractive process, which involves $\pi$ + pomeron
$\rightarrow \pi + \rho$. Both Primakoff production and diffractive
production give events at small t, the four-momentum transfer to the target
nucleus. A simple cut on t is not the best way to make the separation.
Rather, the t-distribution d$\sigma$/dt must be fit in terms of:
$d\sigma/dt=F^2_A(t)(ZC_1/\sqrt t + C_2 exp(at))^2$; where F$_A$(t) is
the nuclear
form factor. The Coulomb Primakoff events follow the C$_1$ term, and the
diffractive pomeron events follow the C$_2$ term. For the pomeron events
only, $d\sigma/dt=F^2_A(t)C_2^2exp(12t) \sim C_2^2$ for t$R^2_A/3<<1$.
Zielinski et al. \cite {ferbel2} did not include an interference term for
three pion production, since Coulomb-produced final states have Gottfried-
Jackson helicity M=$\pm$1, while strong production occurs dominantly with
M=0. We consider here only the coherent Primakoff and diffractive cross
sections, in which the value t is restricted to the coherent region. Zielinski
et al. \cite {ferbel2} define the coherent region for three pion production as
$t<t*$ where $t*=0.4A^{-2/3}GeV^2$ for incident 200 GeV pions. The value t*
corresponds to the expected first minimum of the t-distribution for a target of
nucleon number A. Similar t* definitions will be used for other final
diffractive states.
This work shows how one can attain the required t-resolution needed to
guarantee coherrent diffraction.
For a target nucleus with A=208, the maximum nuclear recoil
energy is $E*=t*/(2 \times 208 \times 0.93)$ = 30 KeV, below the excitation
energy of low lying nuclear levels.

For the diffractive data, following separation from the Coulomb cross section,
one is left with the difficult task of estimating the acceptance and
reconstruction efficiency. The efficiency is sensitive to the polar angle
distribution of the decay pions, which is different for different possible
spin-parity states of the decaying system. A spin-parity analyses is done
\cite{ferbel2,coll}, which is also sensitive to the number of terms in the
model. Usually spins greater than J=3 are ignored; otherwise there are no
convergent fits. This analysis was carried out for the three-pion case, but
would be more difficult for more complex final states. In addition, the
three-pion analysis was restricted to M$_{3\pi}$ values between 0.8 and 1.5
GeV. For larger masses, the f$\pi$ decay mode becomes important, and the
partial wave structure is more complex. The efficiencies were determined in
this way to $\pm 10\%$ for the three-pion case. Similar analyses are required
for each diffractive channel and mass range measured.

We consider diffractive events proceeding through pomeron exchange. For the
$\pi^-$ pomeron $\rightarrow \pi^-\rho$ transition, G-parity is negative
for initial and final states. For the $\pi^-$ pomeron $\rightarrow
\pi^-\omega$ transition, G-parity is negative for the initial state, and
positive for the final state. Therefore, the soft cross section is
dominated by photon exchange for the $\pi\omega$ final state. By studying
both $\pi\rho$ and $\pi\omega$ final states, we can better learn how to
separate Primakoff from diffractive events. The $\omega$ is observed via it
decay to $\pi^+ \pi^- \pi^0$. The invariant mass spectra of the $\pi\omega$
and $\pi\rho$ systems produced are important, and will also be studied.

{\bf A) Inelastic Coherent Diffraction Cross Sections.}

Soft coherent diffractive dissociation of an incident pion by a nuclear
target can provide important experimental tests of the idea of size
fluctuations in the projectile wave function \cite
{fms93b,blat,bbfs,fs,FMS,bfs}. The target remains in its ground state, as
the incident pion diffractively dissociates. The incident pion can be
considered as a superposition of different configurations, having different
sizes. Large inelastic diffractive cross sections arise only if there are
significant differences in the absorption cross sections of the different
configurations, as described in references \cite{fms93b,FMS} and references
therein. For example, the pion wave function can be expanded into states of
$q\bar{q}$, $q\bar{q}g$, $qq\bar{q}\bar{q}$, $qq\bar{q}\bar{q}g$,
$q\bar{q}gg$, etc. Some of these Configurations such as $q\bar{q}$ have
Small Size, some have Normal Size, some have Large Size. These are labelled
as SSC, NSC, LSC. The time scale for fluctuations of the incident pion of
mass m into an excited state of mass M is given by the uncertainty
principle, as $\tau\sim \hbar/(E(m)-E(M))$. For large $p_{lab}$, the energy
denominator $\approx (m^2-M^2)/2 p_{lab}$ is small, and the fluctuation
time is long. The excited state M can move a considerable distance before
decaying, the coherrence length $l_c=2p_{lab}/(M^2-m^2)$, greater than the
diameter of a  target nucleus \cite{fms93b}. The interactions occur between
the excited configuration and target material over the coherrence length,
so that the amplitudes from the entire target for the diffractive
dissociation add coherently and constructively. The incident pion, entering
the nucleus in a specific initial configuration, can be treated as frozen
in that configuration as it passes through the entire nucleus.

FMS \cite{fms93b,FMS} described a two component model of the pion
projectile, to illustrate the basic idea of diffraction as formulated by
Feinberg and Pomeranchuk \cite{fp}, and by Good and Walker \cite{gw}.
The two component wave function is
taken here as:\\ $\mid \pi> =a\mid SSC> + b\mid NSC>$. If the
two components are absorbed with equal strength $\epsilon$, the final state
is just $\epsilon \mid\pi>$, and no inelastic states are produced.
Otherwise, the final state does not coincide with $\mid \pi>$ and inelastic
diffraction takes place \cite{fms93b}. The interaction between a SSC pion
and the target is weak, because color fields in the closely packed SSC
cancel each other. The term {\it color fluctuations}  is used to describe
how the pion fluctuates between its various configurations, and how color
dynamics affects the interaction strengths of the different configurations
\cite{fms93b}. For simplicity, we do not discuss here a number
\cite{fms93b} of possible dynamical mechanisms for the different strengths
of different configurations other than size fluctuations. In this
framework, if the inelastic diffraction cross sections are large, then the
pion wave function must have significant size fluctuations, which is in
line with intuition based on quark models of a hadron. Understanding such
fluctuation effects is simplest at high energies, for which the coherence
length can be significantly larger than the nuclear diameter. But energy
dependent experiments could still be of great value in showing the onset of
the coherrence effects.

High energy diffractive processes have been described \cite{fms93b} in terms of
a probability $P(\sigma)$ that a configuration interacts with a cross section
$\sigma$. $P(\sigma)$ estimated from data is broad; in line with the view that
different size configurations interact with widely varying strengths or cross
sections. One can describe $P(\sigma)$ in terms of its moments: $\langle
\sigma^n\rangle =\int \sigma^n P(\sigma) d\sigma $. The zeroth moment is unity,
by conservation of probability, and the first corresponds to the total
hadron-nucleon cross section $\sigma_{tot}$ (hN). The second moment has been
determined from available diffractive dissociation data. Different
determinations \cite{fms93b, FMS} give consistent values for $\langle
\sigma^2\rangle$, the variance of the distribution: $\omega_{\sigma} \equiv
\left( \langle \sigma^2 \rangle -\langle \sigma \rangle^2 \right) /  \langle
\sigma^2\rangle$ , with $\omega_{\sigma}(p) \sim 0.25$ and
$\omega_{\sigma}(\pi) \sim 0.4$, for incident momenta near 200 GeV/c. The part
of the pion total cross section associated with SSC is roughly 2-5\%, the
integral of P($\sigma$) for $\sigma$ values less than 5 mb \cite{fms93b}.
It has been suggested \cite{jet} to look for the SSC via the hard
dissociation of a $q\bar{q}$ pion to two mini-jets. But it is useful to
consider also whether one can identify SSC contributions in other reaction
channels, including those considered here.

Nuclear hadron inelastic coherent diffractive cross sections provide
experimental tests of size or color fluctuations. The total diffractive cross
section for an incident pion is given as \cite{fms93b}:
\bea &\sigma_{diff}(A)=\int d^2B \times &\nonumber\\ &\left[ \int d\sigma
P(\sigma) \sum_n \left[<\pi\mid F(\sigma,B) \mid n>^2 \right]- \left[\int
d\sigma P(\sigma) <\pi \mid F(\sigma,B) \mid \pi> \right]^2\right].&\qquad
\label{cohdif} \eea Here $ F(\sigma,B) = 1 - e^{-{1\over 2}\sigma T(B)}$,
$T(B)=\int_{-\infty}^\infty \rho_A(B,Z)dz$, with $\rho_A(B,Z)$ the nuclear
density. The direction of the beam is $\hat Z$ and the distance between the
projectile and the nuclear center  is $\vec R=\vec B + Z\hat Z$. Related
equations were given previously within the two-gluon exchange model by
Kopeliovich et al. \cite{bk}; and within the framework of color
transparency by Bertsch et al. \cite {bertsch}.

For the extreme  black disk  limit, the function $F(\sigma,B)$ in Eq. 1 is
unity   for positions inside the nucleus and  zero otherwise, so that
$\sigma_{diff}$ vanishes \cite{fms93b}. One can also show that color
fluctuations lead to non-vanishing diffractive cross sections by
considering the consequences of ignoring the fluctuations of the cross
sections in Eq. 1. In that case, $\ps$ is a delta function, $\ps =\delta
(\sigma - \sigma_{tot}(\pi N))$, and this leads \cite{fms93b} again to
$\sigma_{diff}(A) = 0$.

Such vanishing cross sections contrast strongly with the color fluctuation
calculation of FMS \cite {fms93b} using realistic $\ps$. The FMS
calculation for the total diffractive cross section leads to
$\sigma^{pA}_{diff}(A) \propto A^{0.80}$ for  $A \sim 16$ and $A^{0.4}$
for $A \sim 200$. For the pion projectile, FMS find
$\sigma^{\pi-A}_{diff}(A) \propto A^{1.05}$ for  $A \sim 16$ and $A^{0.65}$
for $A \sim 200$. The FMS calculation agrees well with the A-dependence  of
semi-inclusive data  of \cite {mol} on $n+ A \rightarrow p\pi^- + A$ for
$p_n \sim 300 GeV$ and of \cite{ferbel2} on $\pi^+  + A \rightarrow \pi^+
+\pi^+  +\pi^-  +A$ for $p_{\pi^+}$ = 200 GeV. The major part of the
inelastic diffractive cross section $\sigma$(diff) and its A-dependence
come from configurations centered around those of normal size pions and
neutrons \cite {fms93b}.

 The FMS calculation predicts \cite{fms93b} very large diffractive cross
sections; 40-50 mb for $\pi$-Nucleus and n-Nucleus interactions at 200 GeV
at A $\sim$ 100 and p$_{\pi} \sim$ 200 GeV/c. The specific observed pion
and neutron channels described above correspond only to roughly 20\% and
4\% respectively, of this prediction \cite{fms93b}. A proper demonstration
of the success of the color fluctuation predictions requires experimentally
measuring the cross section and A-dependence of a larger percentage of the
total diffractive cross section. Such future experiments should measure
soft coherent diffractive dissociation of a pion to L-pion final states
with L=3,5,7,9 charged or charged plus neutral pions; mixed pion/kaon final
states with L pions and M kaons, two-meson final states such as $\pi^-\rho$
and $K^{0*}$K$^-$, as well as baryonic states such as $p\bar{p} L \pi$.
Other channels, discussed by Good and Walker \cite{gw}, are
$\pi \rightarrow K \bar{K^0}$,
$\pi \rightarrow K \bar{K^0} \pi^0$,
$\pi \rightarrow \bar{p} n$,
$\pi \rightarrow \bar{\Lambda} \Sigma$.
It
is important to get improved data compared to Zielinski et al. \cite
{ferbel2} for the three-pion case, and new data for the many other
channels. Soft diffractive dissociation cross section data for other
hadrons are also of great interest; as for $\Sigma^- \rightarrow \Lambda
\pi^-$ and other $\Sigma$ channels that can be studied in E781. Such new
experiments will also thereby determine how the total diffractive cross
section is distributed between the different possible channels
\cite{fms93b}. Theoretical prediction are needed for such
channel-distributions; they are not yet available. Such data and
calculations may further clarify our understanding of color fluctuations.

At a given pion beam energy, a simple guess is that the same A-dependence
is expected for all the diffractive channels. It would be surprising and
interesting if a particular semiexclusive reaction would have a very
different A-dependence from that predicted for the total $\sigma$(diff).
That would open the posssibility, not predicted, that such a particular
channel could be more sensitive to SSC or LSC. Different semiexclusive
reactions can be studied in E781. A variety of different targets are needed
to be sensitive to the interesting change
predicted in the A-dependence between light
and heavy targets. The A-dependence for reactions discussed above can be
further subclassified according to the final state invariant and transverse
masses. The reactions should also be studies on the  proton target. Aside
from the extension of the A-dependence to A=1, such data are the natural
input for competing theoretical approaches for hadron-Nucleus interactions,
based on the usual quasi-free approximations.

We discuss now in more detail detecting $\pi^-\rho$ and $K^{0*}$K$^-$
two-meson final states, where the two outgoing mesons have roughly equal
and opposite p$_T$. The $\rho$ and K* are easy to detect, with $\rho
\rightarrow \pi^+ \pi^-$ and $K^{0*} \rightarrow \pi^- K^+$. We have
relative transverse momentum $\Delta(p_T$) for the two outgoing mesons, and
total transverse momentum transfer $\Sigma p_T$ to the target nucleus.
Extremely small momentum transfer $t<< 3/R_A^2$ are needed for coherence to
be valid \cite{fms93b}. For large $\Sigma$p$_T$, the final state nucleus
can be in an excited state, which would destroy coherence. The relative
transverse mass m* is given by m*$^2 \approx 4\Delta^2(p_T$) for a
two-particle final state. Consider experiments such as E781 for which the
incident energy satisfies $E >> 2p_T^2 R_A$ or equivalently $E >> m*^2
R_A/2$, where the transverse mass is m* is 0.6-1.0 GeV and R$_A$ is the
nuclear radius. Using 1 fm=5 (GeV/c)$^{-1}$, with R$_A$=25 (GeV/c)$^{-1}$
for a 5 fm target radius, and m*=1GeV, this condition corresponds to $E >>$
12.5 GeV. The 1 GeV transverse mass condition corresponds to an upper limit
on $\Delta(p_T$) of 1 GeV/c, a soft process.

Each configuration of the pion wave function can contribute to a two-meson
final state, in which each meson has p$_T$ in opposite directions. The
$q\bar{q}$ component can emit the $q$ and $\bar{q}$ with opposite P$_T$,
and then pick up a slow $q\bar{q}$ from the vacuum. For the $q\bar{q}$-glue
component, there are other possibilities. The $q$ and $\bar{q}$ can be
emitted with opposite P$_T$, together with a soft gluon. This gluon can
itself dissociate to $q\bar{q}$, which can join the original $q\bar{q}$ to
form the final meson$_1$ and meson$_2$.

 Different channels can certainly have different relative contributions
from different pion configurations. Consider the $d\bar{u}$ component of
the pion as an example. Studies of the structure function of the kaon show
that the s-quark carries the larger fraction of the kaon momentum. The
$d\bar{u}$ of the $\pi^-$ separate, and say combine with an $s\bar{s}$ from the
vacuum, to form $d\bar{s}$ $K^{0*}$ and $s\bar{u}$ (K$^-$). The overlap with
the final kaon wave functions may be very small; since the produced $s\bar{s}$
pair is slow, but the strange quarks in the final state kaons should be
leading. This is not the case for the $q\bar{q}g$ component. Here, the momentum
fraction x(g) of the gluon may be large, the gluon may dissociate to
$s\bar{s}$, and the resulting s quarks may be leading. For $\pi^- \rightarrow
\pi^- \rho$, the $d\bar{u}$ of the $\pi^-$ separate, and say combine with a
$u\bar{u}$ from the vacuum, to form $d\bar{u}$ ($\pi^-$) and $\rho^0$
($u\bar{u}$). In this case, there may be better overlap with the final state
$\pi$ and $\rho$ wavefunctions. The $u$ and $\bar{u}$ do not have to be leading
in the final mesons. This example illustrates that different configurations of
the pion, say $d\bar{u}$, can contribute differently to different two-meson
channels. It is possible therefore that some particular channel (or mass
region) may have a different A-dependence than others, or may be more or less
sensitive to SSC.

The total soft diffractive cross section probability for a pion can be
approximated by \cite{fms93b}:
\bea
\sigma_{diff}^{appr}(A) =
(\omega_{\sigma} \langle \sigma^2 \rangle /4)
\int d^2B T^2(B) exp[-\sigma_{tot}(h N) T(B)].
\eea
This expression is found \cite{fms93b} starting with Eq. 1, and expanding
P($\sigma$) in a Taylor series around $\bar{\sigma}=\sigma_{tot}(h N)$.
This formula in the framework of the color fluctuations calculation, gives
a clear relationship between the measured total diffractive cross section
for target A, and the total hadron-Nucleon cross section. A confirmation of
this relationship would provide another test for the color fluctuation
framework. The color fluctuation framework also gives predictions
\cite{fms93b} for the zero-degree total diffractive differential cross
sections, which can provide yet further checks.

The 1-3 GeV higher transverse mass range for two-meson diffractive
transitions involves a mixture of soft and hard processes. Theoretical
calculations are difficult for this mixed region. It is difficult to get
high quality data for transverse mass values significantly higher than 1
GeV, considering that the cross sections and statistics will be too small.
Some low statistics data for this mass range may nonetheless motivate
theoretical efforts. One may expect \cite{kop} also that two-meson data at
higher transverse mass would be more sensitive to the SSC, as was
demonstrated recently for two-jet \cite{jet} diffraction.
A natural guess for fixed invariant mass M is that the
transparency and A-dependence will increase with increasing transverse mass
\cite {kop,fms93b}.
One may also expect
\cite{bertsch} that diffractive production of charm may be sensitive to SSC.

{\bf B) Radiative Widths of Mesons.}

   High energy pion experiments at FNAL and CERN can obtain new
high statistics data for radiative transitions leading from the pion to the
$\rho$, to the a$_1$(1260), and to the a$_2$(1320). These radiative
transition widths are predicted by vector dominance and quark models. They
were studied in the past by different groups, but independent data would
still be of value. For $\rho \rightarrow \pi \gamma$, the widths obtained
\cite{jens,ziel2,hust,capr} range from 60. to 81. KeV. For a$_1$(1260)
$\rightarrow \pi \gamma$, the width given \cite{ziel} is $\Gamma = 640. \pm
246.$ KeV; for a$_2$(1320) $\rightarrow \pi \gamma$, the width given
\cite{ciha} is $\Gamma = 295 \pm$ 60 KeV; and for b$_1$(1235), the width given
\cite{coll} is $\Gamma = 230 \pm$ 60 KeV. Radiative transitions of $\Sigma
\rightarrow \Sigma*$ can also be studied \cite{moin1,lipk}. The a$_1$(1260)
radiative width is related to the pion polarizability \cite {moin1,hols,xsb}.

Studies are also possible for Primakoff production of exotic mesons.
Consider the search for C(1480) 1$^{--}$ via Primakoff production;
described by L. Landsberg \cite{c1480}. It involves Primakoff production of
C(1480) with a $\pi^-$ beam at 600 GeV, and observation of the C decay by:\\
a)  C $\rightarrow \phi \pi^-$ (with $\phi \rightarrow K^- K^+$)\\
b)  C $\rightarrow \omega \pi^-$ (with $\omega \rightarrow  \pi^+ \pi^- \pi^0$;
and $\pi^0 \rightarrow \gamma \gamma$).\\
c)  C $\rightarrow \omega \pi^-$ (with $\omega \rightarrow \pi^0 \gamma$ ; and
$\pi^0 \rightarrow \gamma \gamma$).\\
Consider also the search for 1$^{-+}$ Hybrid Mesons, as described by M.
Zielinsky et al. \cite{hybrid}. One can look for Primakoff production
with a 600 GeV $\pi^-$ beam of the Hybrid meson HY, via observation of a number
of decay channels: \\
a) HY $\rightarrow \pi^- f1(1285)$ (f1(1285) $\rightarrow \pi^+\pi^-\eta; \eta
\rightarrow \gamma \gamma$)\\
b) HY $\rightarrow \pi^- \eta (\eta \rightarrow  \gamma \gamma$)\\
c) HY $\rightarrow \rho^0 \pi^- (\rho^0 \rightarrow \pi^+ \pi^-$)\\
d) HY $\rightarrow \eta' \pi^- (\eta' \rightarrow \pi^+ \pi^- \eta;
\eta \rightarrow \gamma \gamma$).\\

{\bf C) Experimental determination of the $\pi + \rho \rightarrow  \pi +
\gamma$ reaction rate.}

 The gamma production reaction can be studied (via the inverse reaction,
with detailed balance) via the Primakoff reaction $\pi^- + \gamma
\rightarrow \pi^- + \rho^0$. This production rate enters the consideration
of the expected gamma-ray background from the hot hadronic gas phase in
heavy ion collisions, and is important for quark gluon plasma experimental
searches via gamma ray production. Of interest also is the invariant mass
of the produced $\pi\rho$ system. The invariant mass reveals information
regarding the reaction mechanism. For the case of $\pi\rho$ detection, one
may expect the invariant mass to show a spectrum of resonances that have a
$\pi\rho$ decay branch. These include the a$_1$(1260), the $\pi$(1300),
a$_2$(1320), a$_1$(1550), etc. The mass spectrum may also show a high mass
tail region above these resonances. One measures the reaction rate at normal
temperatures for normal mass pion and rho, and intermediate resonances.

Xiong, Shuryak, Brown (XSB) \cite{xsb} calculate photon production (above
0.7 gev) via the reaction $\pi^- + \rho \rightarrow \pi^- + \gamma$. The
reaction $\rho + \pi^- \rightarrow \gamma + \pi^-$ proceeds (according to
XSB) through the a$_1$(1260). XSB estimate Radiative Width (a$_1
\rightarrow \pi \gamma$) = 1.4 MeV, more than two times higher than the
experimental value of Zielinski et al. \cite{ziel}. With this estimated
width, they calculate the high energy photon production cross section. They
include high temperature effects for a hot hadronic gas. A more recent
photon production calculation also involving a$_1$ resonance effects was
given by Song \cite{song}. There are many other theoretical studies for
gamma rays from hadronic gas and QGP in the Quark Matter conferences, and
elsewhere. Some relevant articles are by  Ruuskanen \cite{ruu},  Kapusta et
al. \cite{kapuqm},  Alam et al. \cite{ala}, Nadeau \cite{nad}, and
Schukraft \cite{sch}. One can measure such cross sections for normal mass
mesons, and therefore to experimentally provide the data base for
evaluations of  the utility of gamma production experiments in QGP
searches. One can experimentally check the a$_1$(1260) dominance assumption
of XSB.

In a hadronic gas at high temperature, the $\pi\rho$ interaction can be
near the a$_1$ resonance \cite{xsb}. One must consider also that certain
properties (masses, sizes, parity mixing) of the $\pi$ and $\rho$ and a$_1$
change \cite{xsb,song,dey,asa,her}, and that their numbers increase due to the
Boltzmann factor. XSB expect an increased yield from the hot hadronic gas,
higher than estimated previously by Kapusta et al. \cite{kapu}. One expects
many gamma rays from QGP processes, such as a quark-antiquark annihilation $q
\bar{q}\rightarrow g \gamma$ or Compton processes such as $q g \rightarrow q
\gamma$ and $\bar{q} g \rightarrow \bar{q} \gamma$. Chakrabarty et al.
\cite{chak} studied the expected gamma ray yields from hot hadronic gases and
the QGP. They suggested that gamma rays between 2-3 GeV from the QGP outshine
those from the hot hadronic gas phase.

{\bf D) Investigation of Chiral Anomalies. }

Chiral anomalies can be studied in E781. Before giving experimental details, we
first describe chiral anomalies for a massless field theory, following K. Huang
\cite{huang}. The lagrangian density in such a theory is invariant under a
chiral transformation, which implies that a conserved axial-vector current must
exist. One may ask if this axial current is the gauge-invariant chiral current.
In that case, the divergence of the chiral current should be zero. However, the
correct analysis must account for the fact that the currents are singular
operators. One can then show that the divergence of the chiral current is equal
to an "axial anomaly" term rather than zero. The required conserved
axial-vector current does exist and can be defined; but it is not the chiral
current, it is not gauge invariant, and it does not couple to physical fields.

For the $\gamma$-$\pi$ interaction at low energy, chiral perturbation
theory ($\chi$PT) provides a rigorous way to make predictions; because it
stems directly from QCD and relies only on the solid assumptions of
spontaneously broken SU(3)$_L$ $\times$ SU(3)$_R$ chiral symmetry, Lorentz
invariance and low momentum transfer. Unitarity is achieved by adding pion
loop corrections to lowest order, and the resulting infinite divergences
are absorbed into physical (renormalized) coupling constants
\cite{gl,donn1}. With a perturbative expansion of the effective Lagrangian
limited to terms quartic in the momenta and quark masses (O(p$^4$)), the
method successfully describes many physical processes. At O(p$^4$) level,
the lagrangian includes Wess-Zumino-Witten (WZW) terms \cite{wzw}, which
incorporate the chiral anomalies of QCD. These modify the Ward identities
\cite{gl,ward} for the currents, and also lead to anomalous terms
\cite{huang,gl,wzw,anti} in the divergence equations of the currents. These
anomalies at the O(p$^4$) level lead directly to an interesting
relationship \cite{ter} between the processes $\pi^0 \rightarrow  2 \gamma$
and $\gamma  \rightarrow  3 \pi$. The latter two processes are described by
the coupling constants F$_{\pi}$ and F$_{3\pi}$, respectively. The
F$_{\pi}$ vertex was first described by Adler, Bell, and Jackiw \cite{abj}.
The relationship is \cite{ter}:
\bea
F_{3\pi} = F_{\pi}/(e f^2),
\eea
where e=$\sqrt(4 \pi \alpha)$ and f is the charged pion decay constant.
The experimental confirmation of Eq. 3 would
demonstrate that the O(p$^4$) terms are sufficient to describe F$_{3\pi}$
within the framework of the chiral anomalies.

   For the chiral anomaly, the $\gamma\pi \rightarrow \pi \pi$
reaction was measured
\cite{anti} with 40 GeV pions at Serpukhov via
pion pair production by a pion in
the nuclear Coulomb field ($\pi^{-}$  +  Z $\rightarrow$  $\pi^{-}$  + Z  +
$\pi^0$) ; where the incident pion
interacts with a virtual photon in the
Coulomb field of a nucleus of atomic number Z; and the two final state
pions (typically 20 GeV each) were detected in coincidence.
This reaction is equivalent to  the $\gamma$ +
$\pi^{-}$ $\rightarrow$  $\pi^0$ + $\pi^{-}$ reaction for a laboratory gamma
ray of several hundred MeV incident on a target $\pi^{-}$ at rest.
In the incident pion rest frame, the nucleus Z represents a beam or
cloud of virtual photons sweeping past the pion.
Such a reaction is an
example of the well tested Primakoff formalism \cite{jens,ziel2} that relates
processes involving real photon interactions to production cross sections
involving the exchange of virtual photons.

In the Serpukhov experiment, it was shown
\cite{anti} that the Coulomb amplitude
clearly dominates and yields sharp peaks in t-distributions at very small four
momentum transfers to the target nucleus. The cross
sections corresponding to the sharp peaks in the t-distributions for targets
with different atomic number Z scaled as $Z^2$, further demonstrating the
correspondence with the Primakoff formalism. Background from strong
processes (meson or pomeron exchange)
has an exponential falloff with increasing t.
The Coulomb
cross section is about 0.06 $\mu$barns, for a C$^{12}$ target.

      To illustrate the kinematics, consider the reaction:
%\begin{eqnarray}
$$
{\pi} + Z \rightarrow {\pi}' + Z' + {\pi^0}' \eqno(4)
$$
%\end{eqnarray}
for a 600 GeV incident pion, where Z is the nuclear charge. The
4-momentum of each particle is $P_{\pi}$, $P_Z$, $P_{{\pi}'}$, $P_{Z'}$,
$P_{\pi^0}$,
respectively. In the one photon exchange domain, eqn. 4 is equivalent to:
$$
%\begin{eqnarray}
\gamma + {\pi}  \rightarrow  {\pi}' + {\pi^0}, \eqno(5)
$$
%\end{eqnarray}
and the 4-momentum of the incident virtual photon is k = $P_Z$-$P_{Z'}$.
The cross section for the reaction of eqn. 4
depends on $F_{3\pi}^2$, and on s, t, t$_1$, t$_0$. Here
t is the square of the
four-momentum transfer to the nucleus, $\sqrt{s}$ is the invariant mass of the
$\pi \pi$ final state, t$_1$ is the square of the 4-momentum transfer
between initial and final $\pi^-$ in Eq. 5,
and $t_0$ is the minimum value of t to produce a
mass $\sqrt{s}$.

The data  yield F$_{3\pi}=12.9
\pm 0.9 (stat) \pm 0.5 (sys) GeV^{-3}$, from Antipov et al. \cite{anti}.
The uncertainties do not include aproximately 10\% uncertainties \cite
{anti} in extrapolating F$_{3\pi}$ to threshold (s, t$_1$ approaching
zero), which is where Eq. 3 is strictly valid. In addition, Antipov et al.
use f$_{\pi}= 90 \pm$ 5 MeV, and give the theoretical expectation as
F$_{3\pi}= 10.5 \pm 1.5 GeV^{-3}$. Comparing experiment and theory,
considering the quoted errors, Antipov et al. claimed that the hypothesis
of chiral anomalies and color-SU(3) quark symmetry are confirmed. In fact,
a recent determination by Holstein \cite{hols1} of the pion decay constant
gave a value of 92.4 $\pm$ 0.2 MeV, somewhat lower than the value cited by
the Particle Data Group \cite{pdg} of 93.2 $\pm$ 0.1 MeV. Holstein claims
that the PDG value 93.2 is too large due to incomplete inclusion of
radiative corrections in its extraction. The Holstein value was confirmed
independently by Marciano and Sirlin \cite {mar}. Both the Holstein and PDG
values and errors are significantly different from the value 90 $\pm$ 5 MeV
used by Antipov et al. In what follows, we use the value f= 92.4 $\pm$ 0.2
MeV rather than the 1990 and 1992 PDG value; as this appears to be very
well founded \cite {hols1,mar}, and also leads to more conservative
conclusions. The consequently revised O(p$^4$) expectation for F$_{3\pi}$
is therefore 7.4\%  lower than given by Antipov et al., and also the
uncertainty associated with f is reduced, leading to F$_{3\pi}= 9.72 \pm
0.06 GeV^{-3}$. In that case, the experimental  result of Antipov et al. in
fact differs with the O(p$^4$) chiral anomaly expectation by at least two
standard deviations. A related reaction \cite{amen} to determine F$_{3\pi}$
is $\pi^- + e \rightarrow \pi^- + \pi^0 +e'$, for which an incident high
energy pion scatters inelastically from a target electron in an atomic
orbit. The data uncertainties \cite{amen} in this case are roughly 25\%,
and there are also additional theoretical uncertainties in the
extrapolation to zero momentum transfer. Therefore, the hypothesis of
chiral anomalies at O(p$^4$) is not confirmed by the available $\gamma
\rightarrow 3\pi$ data.

Bijnens et al. \cite{bij1,bij2,bij3} within $\chi$PT studied yet higher
order corrections in the abnormal intrinsic parity (anomalous) sector.
They included one-loop diagrams involving one vertex from the WZW term, and
tree diagrams from the O(p$^6$) lagrangian. Some one-loop diagrams give
finite contributions. Others lead to divergences that are eliminated
by the O(p$^6$) terms. These higher order corrections are small for F$_{\pi}$.
For the F$_{3\pi}$ vertex, they
increase the lowest order value of F$_{3\pi}$, from Eq. 3,
from 7\% to 12\%.
Bijnens et al. give the eq. 3 value of F$_{3\pi}$
as 9.5 GeV$^{-3}$, which corresponds to using the PDG value f=93.2 MeV.
The one-loop and  O(p$^6$) corrections
to F$_{3\pi}$ are comparable in
strength. The loop corrections to F$_{3\pi}$ are not constant over the
whole phase space, due to dependences on the momenta of the 3 pions.
The average effect is roughly 10\%, which then changes the theoretical
prediction
from Eq. 3 to roughly 11. GeV$^{-3}$. As discussed by Bijnens et al.,
the higher order corrections improve the agreement between
theoretical predictions and the data. The large experimental errors however
do not allow one to disentangle the loop effects from the O(p$^6$) effects.
The calculations of Bijnens et al motivate an improved experiment.

The experiment at 40 GeV suffered from the need to disentangle Primakoff
and strong contributions. This problem was a major factor in setting the
systematic uncertainty of the experiment. In E781, at the 600 GeV higher
energy, the strong contribution is negligible, which should significantly
reduce the systematic uncertainty. The 1 MHz pion flux at FNAL E781
will enable superb statistics for a new measurement. Also, the
extrapolation to threshold can be accomplished with significantly smaller
error in future experiments. This is done
for a high statistics experiment by restricting the data set to
significantly lower values of $\sqrt(s)$
and t$_1$, compared to Antipov et
al. One can therefore get improved data for a significantly improved test
of chiral anomalies, with E781. One can test how well $\chi$PT works in
the anomalous sector. How anomalous is the real world anyhow?

\centerline{\bf {CONCLUSIONS}}
This completes the discussion of points A-D. There are experimental
possibilities at FNAL E781, and elsewhere. The cross sections need to be
studied experimentally versus A, incident energy, incident particle type,
$\Delta p_T$, $\Sigma p_T$, invariant mass and transverse mass of the produced
two-meson or multi-pion final systems. Theoretical calculations are available
for many of the reactions discussed, and interesting E781 data should motivate
further theoretical developments.

\centerline{\bf ACKNOWLEDGEMENTS}
Discussions and correspondence
on different aspects of these subjects with G. Brown, J. Dey,
J. Eisenberg, T. Ferbel, L. Frankfurt, S. Gerzon, B. Holstein, B.
Kopeliovich, L. Landsberg, H. Leutwyler,
H. Lipkin, S. Nussinov, E. Piasetzky, J. Russ,
V. Steiner, and  M. Strikman are acknowledged. This work was supported by
the U.S.-Israel Binational Science Foundation (BSF), Jerusalem, Israel.

\end{document}